\documentclass[10pt,final]{elsart}
\usepackage{hyperref}
\usepackage{graphicx}
\usepackage{amsmath,amsfonts}

\newcommand{\f}{\frac}

\newcommand{\m}{\mathbf}

\newcommand{\bs}{\boldsymbol}
\begin{document}
\begin{frontmatter}

\title{Hydrodynamic Model for the System of Self Propelling Particles with Conservative Kinematic
Constraints; Two dimensional stationary solutions}
\author[ned]{V.I. Ratushnaya}
\author[odes]{V.L. Kulinskii}\ead{kulinskij@onu.edu.ua}
\author[eng]{A.V. Zvelindovsky}
\author[ned]{D. Bedeaux}
\address[ned]{Colloid and Interface Science group, LIC, Leiden University, \\
P.O. Box 9502, 2300 RA Leiden, The Netherlands}
\address[odes]{Department for Theoretical Physics, Odessa National University, \\
Dvoryanskaya 2, 65026 Odessa, Ukraine}
\address[eng]{Department of Physics, Astronomy \& Mathematics, University of \\
Central Lancashire, Preston PR1 2HE, United Kingdom}

\begin{abstract}
In a first paper we proposed a continuum model for the dynamics of
systems of self propelling particles with kinematic constraints on
the velocities. The model aims to be analogous to a discrete
algorithm used in works by T. Vicsek et al. In this paper we prove
that the only types of the stationary planar solutions in the
model are either of translational or axial symmetry of the flow.
Within the proposed model we differentiate between finite and
infinite flocking behavior by the finiteness of the kinetic energy
functional.
\end{abstract}
\end{frontmatter}

\section{Introduction}
The dynamics of systems of particles subjected to nonpotential
interactions remains poorly understood. The absence of a
Hamiltonian for such systems, which generally are far from
equilibrium, hampers applying the machinery of statistical
mechanics based on the Liouville equation. Many attempts have been
made to investigate these systems using discrete algorithms to
model this behavior. In nature there are many examples of such
systems \cite{parishbook}. Since the discrete algorithms are hard
to describe analytically it is natural also to consider continuum
models of a hydrodynamic type. In standard hydrodynamics the
relation between microscopic kinetics (Boltzmann-type equations)
and Navier-Stokes equations is a standard topic of research
\cite{succi2001}. For the systems of interest the construction of
corresponding kinetic equations based on the specific dynamic
rules and their connection with the hydrodynamics equations seems
to be unknown so far and is worth studying. One can expect that
the continuous description of the collective behavior like
swarming and flocking leads to quite unusual hydrodynamics.

In our first paper \cite{usepll2005} we proposed a hydrodynamic
model which can be considered to be the continuum analogue of the
discrete dynamic automaton proposed by Vicsek et al.
\cite{cvaprl1995} for a system of self propelling particles. It
uses the continuity equation
\begin{equation}\label{conteq}
\frac{\partial n(\mathbf{r},t)}{\partial t}+\mathop{\rm
div}\nolimits\left(
n(\mathbf{r},t)\,\mathbf{v}(\mathbf{r},t)\right) =0\,,
\end{equation}
which implies that the total number of particles
\begin{equation}\label{totnum}
    N = \int n(\m{r},t) \,d\m{r}
\end{equation}
is constant. The kinetic energy of a co-moving volume
\begin{equation}\label{kin}
T = \f{1}{2}\int n(\m{r},t) \m{v}^2(\m{r},t) d\m{r}
\end{equation}
is also conserved
\begin{equation}\label{totenrg}
    \frac{d}{dt}T =0\,.
\end{equation}
Using Eq.~\eqref{conteq} and Eq.~(\ref{totenrg}) it can be shown
that a field $\bs{\omega}$ exists such that the Eulerian velocity,
$\mathbf{v}(\mathbf{r},t),$ satisfies:
\begin{equation}
\frac{d}{dt}\mathbf{v}\left( \mathbf{r},t\right) =\boldsymbol{\omega}\left(
\mathbf{r},t\right) \times \mathbf{v}\left( \mathbf{r},t\right) .
\label{1.3}
\end{equation}
This equation can be considered as the continuous analogue of the
conservative dynamic rule used by Vicsek et al. \cite{cvaprl1995}.

We proposed the following ``minimal`` model for the field of the
angular velocity $\boldsymbol{\omega}\left(\mathbf{r},t\right)$
which is linear in spatial gradients of the fields $n(\m{r},t)$ or
$\m{v}(\m{r},t)$:
\begin{equation}
\mathbf{\omega }\left( \mathbf{r},t\right) =\int K_{1}\left( \mathbf{r}-%
\mathbf{r^{\prime }}\right) \,n(\mathbf{r^{\prime }},t)\mathop{\rm rot}%
\nolimits\mathbf{v}(\mathbf{r^{\prime }},t)\,d\mathbf{r^{\prime
}+}\int
K_{2}\left( \mathbf{r}-\mathbf{r^{\prime }}\right) \nabla n\left( \mathbf{%
r^{\prime }},t\right) \times \mathbf{v}(\mathbf{r^{\prime }},t)d\mathbf{%
r^{\prime }}\,.  \label{1.4}
\end{equation}
$\bs{\omega}$ has the proper pseudovector character. The averaging kernels $%
K_{1}\left(\mathbf{r}-\mathbf{r^{\prime}}\right)$ and $K_{2}\left(\mathbf{r}-%
\mathbf{r^{\prime}}\right)$ should naturally decrease with the distance in
realistic models. They sample the density and the velocity around $\mathbf{r}$%
\ in order to determine
$\boldsymbol{\omega}\left(\mathbf{r},t\right)$. In the first paper
we concentrated on $K_{1}$. The detailed derivation of the above
equations from the discrete models based on the automaton proposed
by Vicsek et al. \cite{cvaprl1995,cvbenjacohenpre1996} will be the
subject of a future paper.

Note that the models based on Eqs.~\eqref{conteq}-\eqref{1.4}
allow solutions of uniform motion in the form of a solitary
packet:
\begin{equation}
n(\mathbf{r},t)=n_{0}(\mathbf{r}-\mathbf{v}_{0}\,t)  \label{1.5}
\end{equation}
with $\mathbf{v}_{0}$ independent of position and time. The contribution to $%
\boldsymbol{\omega }$ due to $K_{1}$\ is zero for an arbitrary density
distribution $n_{0}$. The contribution due to $K_{2}$\ is zero for density
distributions $n_{0}$ which only depend on the position in the $\mathbf{v}%
_{0}$\ direction. In this second case it follows from the
continuity equation that $n_{0}$\ should be everywhere constant.
The density distribution $n_{0}$ should be chosen such that the
number of particles, and, correspondingly, the total kinetic
energy are finite. The solutions of such type also were found
analytically in \cite{topaz/siam/2004} and observed in simulations
\cite{gregchate/prl/2004}. Note that such solutions exist not only
in nonlocal case like in \cite{topaz/siam/2004} but also for the
local model which we consider below.

Within the first order of perturbation theory on small deviation
of density and velocity fields the solitary solution given by
Eq.~\eqref{1.5} shows neutral stability; i.e. the perturbations
grows linearly for small $t$.

We restrict our discussion to the simple case of averaging
kernels, which are $\delta $-functions:
\begin{equation}
K_{j}\,(\mathbf{r}-\mathbf{r}^{\prime })=s_{j}\,\delta (\mathbf{r}-\mathbf{r}%
^{\prime })\,,\quad \text{\ where \ \ }j=1\text{ or 2}.
\label{1.6}
\end{equation}
We will call this the local hydrodynamic model (LHM). In the first
paper, where we only
considered $K_{1}$, we scaled $K_{1}$\ by dividing by $\left| s_{1}\right| $%
\ and the density $n$ by multiplying with $\left| s_{1}\right| $. This made
it then possible to restrict the discussion to $s_{1}$ is plus or minus one.
The disadvantage of this scaling procedure is that it changes the
dimensionality of $K_{j}$ and $n$. For two kernels it becomes impractical.
We note that $s_{j}$\ is given by:
\begin{equation}
s_{j}=\int \ K_{j}\left( \mathbf{r}\right) d\mathbf{r}.  \label{1.7}
\end{equation}
For the local model Eq.(\ref{1.4}) reduces to:
\begin{equation}
\boldsymbol{\omega }\left( \mathbf{r},t\right) =s_{1}\,n(\mathbf{r},t)%
\mathop{\rm rot}\nolimits\mathbf{v}(\mathbf{r},t)\text{ }\mathbf{+}\text{ }%
s_{2}\mathbf{\nabla }n\left( \mathbf{r},t\right) \times \mathbf{v}(\mathbf{r}%
,t).  \label{lhm}
\end{equation}
and Eq.~\eqref{1.3} for the velocity becomes
\begin{equation}
\frac{d}{dt}\mathbf{v}\left( \mathbf{r},t\right) =s_{1}\,n(\mathbf{r},t)%
\mathop{\rm rot}\nolimits\mathbf{v}(\mathbf{r},t)\times \mathbf{v}\left(
\mathbf{r},t\right) +s_{2}\left( \mathbf{\nabla }n\left( \mathbf{r},t\right)
\times \mathbf{v}(\mathbf{r},t)\right) \times \mathbf{v}\left( \mathbf{r}%
,t\right)\,.  \label{lhmv}
\end{equation}
Note that the second term on the right-hand-side of
Eq.~\eqref{lhmv} corresponds to the rotor chemotaxis force when
the number density is proportional to the attractant density
introduced in \cite{cvbenjacohenpre1996}.

In the following section we will show that the only stationary
solutions for LHM with $\boldsymbol{\omega}$ field given by
Eq.~\eqref{lhm} are either the solutions of uniform motion (see
Eq.~\eqref{1.5}) or the radially symmetric planar solution which
will be considered in detail in the following section.

In the second section we investigate the properties of the
stationary radially symmetric solutions of the local hydrodynamic
model for some special cases. Conclusions are given in the last
section.
\section{Possible types of stationary states for the local hydrodynamical
model}
The equations of motion to be solved are Eqs.~\eqref{conteq} and
~\eqref{lhmv}. In order to find a class of 2D stationary solutions
we consider this problem in a generalized curvilinear orthogonal
coordinate system $(u,v)$, which can be obtained from the
Cartesian one $(x,y)$ by some conformal transformation of the
following form:
\begin{equation}
u+iv=F\left(z\right),  \label{3.2}
\end{equation}
where $F(z)$ is an arbitrary analytical function of $z = x + i y$.
In a curvilinear orthogonal coordinate system the fundamental
tensor has a diagonal form, $g_{ik}=g_{ii}\delta _{ik}$, where the indices $%
i,j$ are either $u$ or $v$. The square of linear element in
conformal coordinates is
\begin{equation}
d\mathbf{s}^{2}=\frac{1}{D(u,v)}\left( du^{2}+dv^{2}\right) ,
\label{3.3}
\end{equation}
where
\begin{equation}
D(u,v)=\frac{\partial \left( u,v\right) }{\partial \left( x,y\right) }%
=\left( \frac{\partial u}{\partial x}\right) ^{2}+\left( \frac{\partial u}{%
\partial y}\right) ^{2}=\left( \frac{\partial v}{\partial x}\right)
^{2}+\left( \frac{\partial v}{\partial y}\right)
^{2}=\frac{1}{\sqrt{g}} \label{3.4}
\end{equation}
is the Jacobian of the inverse conformal transformation from the
arbitrary curvilinear orthogonal to Cartesian coordinates,
$(u,v)\rightarrow (x,y)$. Furthermore $g=g_{uu}g_{vv}$ is the
determinant of the metric tensor. For a conformal transformation
$g_{vv}=g_{uu}=1/D$.

The differential operations are given by the following expressions
\cite{madelung}:
\begin{eqnarray}
\nabla \phi &=&\sqrt{D}\left( \mathbf{e}_{u}\frac{\partial \phi }{\partial u}%
+\mathbf{e}_{v}\frac{\partial \phi }{\partial v}\right),  \label{3.5a} \\
\mathop{\rm div}\nolimits\mathbf{a} &=&D\left[ \frac{\partial }{\partial u}%
\left( \frac{a_{u}}{\sqrt{D}}\right) +\frac{\partial }{\partial
v}\left(
\frac{a_{v}}{\sqrt{D}}\right) \right],  \label{3.5b} \\
\mathop{\rm rot}\nolimits\mathbf{a} &=&D\left[ \frac{\partial }{\partial u}%
\left( \frac{a_{v}}{\sqrt{D}}\right) -\frac{\partial }{\partial
v}\left(
\frac{a_{u}}{\sqrt{D}}\right) \right], \label{3.5c} \\
\Delta \phi &=&D\left[ \frac{\partial ^{2}\phi }{\partial u^{2}}+\frac{%
\partial ^{2}\phi }{\partial v^{2}}\right]\,.  \label{3.5d}
\end{eqnarray}
Here $\mathbf{e}_{u}$ and $\mathbf{e}_{v}$\ are orthonormal base
vectors in the directions of increasing $u$ and $v$ respectively.
These base vectors are functions of the coordinates $u$ and $v$.
The projections of the vectorfield $\mathbf{a}$\
on these directions are $a_{u}=\mathbf{a\cdot e}_{u}$\ and $a_{v}=\mathbf{%
a\cdot e}_{v}$. Using Eqs.~\eqref{3.5a}-\eqref{3.5d} for the
velocity field given by $\mathbf{v}=
\mathrm{v}_{u}\,\mathbf{e}_{u}+\mathrm{v}_{v}\,\mathbf{e} _{v}$,
 one obtains:
\begin{equation}
\left(\mathbf{v}\cdot\nabla\right)\,\mathbf{v}=\sqrt{D}\left[ \mathrm{v}_{u}%
\frac{\partial (\mathrm{v}_{u}\mathbf{e}_{u})}{\partial u}+\mathrm{v}_{v}%
\frac{\partial (\mathrm{v}_{u}\mathbf{e}_{u})}{\partial v}+\mathrm{v}_{u}%
\frac{\partial (\mathrm{v}_{v}\mathbf{e}_{v})}{\partial u}+\mathrm{v}_{v}%
\frac{\partial (\mathrm{v}_{v}\mathbf{e}_{v})}{\partial v}\right],
\label{3.6a}
\end{equation}
\begin{equation}
\mathop{\rm rot}\nolimits\mathbf{v}\times \mathbf{v}=D\left[
\frac{\partial
}{\partial u}\left( \frac{\mathrm{v}_{v}}{\sqrt{D}}\right) -\frac{\partial }{%
\partial v}\left( \frac{\mathrm{v}_{u}}{\sqrt{D}}\right) \right] (-\mathrm{v}%
_{v}\mathbf{e}_{u}+\mathrm{v}_{u}\mathbf{e}_{v}),  \label{3.6b}
\end{equation}
\begin{equation}
\left( \mathbf{\nabla }n\times \mathbf{v}\right) \times \mathbf{v=}\sqrt{D}%
\left[ \frac{\partial n}{\partial u}\mathrm{v}_{v}-\frac{\partial n}{%
\partial v}\mathrm{v}_{u}\right] \left( -\mathrm{v}_{v}\mathbf{e}_{u}+%
\mathrm{v}_{u}\mathbf{e}_{v}\right),  \label{3.6c}
\end{equation}
\begin{equation}
\mathop{\rm div}\nolimits(n\mathbf{v})=D\left[ \frac{\partial }{\partial u}%
\left( \frac{n\mathrm{v}_{u}}{\sqrt{D}}\right) +\frac{\partial }{\partial v}%
\left( \frac{n\mathrm{v}_{v}}{\sqrt{D}}\right) \right] =0.
\label{3.6d}
\end{equation}

Substituting Eqs.~\eqref{3.6a}-\eqref{3.6d} into
Eqs.~\eqref{conteq} and \eqref{lhmv} we obtain the following
system of equations, which determines all possible stationary
flows for the LHM:
\begin{equation}\label{vu1}
\mathrm{v}_{u}\frac{\partial\mathrm{v}_{u}}{\partial u}+(1-s_{1}n)\mathrm{v}%
_{v}\frac{\partial\mathrm{v}_{u}}{\partial
v}+%
\mathrm{v}_{u}\mathrm{v}_{v}\left(
f_{3}+\frac{s_{1}n}{2}\frac{\partial \ln D}{\partial v} -
s_2\,\f{\partial n}{\partial v}\right) + \mathrm{v}_{v}^{2}\left(
f_{4}-\frac{s_{1}n}{2}\frac{\partial \ln D}{\partial u} +
s_2\,\f{\partial n}{\partial
u}\right)+s_{1}n\mathrm{v}_{v}\frac{\partial\mathrm{v}_{v}}{
\partial u}=0,
\end{equation}
\begin{equation}\label{vv1}
\mathrm{v}_{v}\frac{\partial\mathrm{v}_{v}}{\partial v}+(1-s_{1}n)\mathrm{v}%
_{u}\frac{\partial \mathrm{v}_{v}}{\partial u}+\mathrm{v}_{u}\mathrm{v}%
_{v}\left(f_{2}+\frac{s_{1}n}{2}\frac{\partial\ln D}{\partial
u}-s_2\,\f{\partial n}{\partial u}\right) +%
\mathrm{v}_{u}^{2}\left( f_{1}-\frac{s_{1}n}{2}\frac{\partial\ln
D}{\partial v} +s_2\,\f{\partial n}{\partial v}\right)
+s_{1}n\mathrm{v}_{u}\frac{\partial\mathrm{v}_{u}}{\partial v}=0,
\end{equation}

\begin{equation}
\frac{\partial}{\partial u}\left(\frac{n\mathrm{v}_{u}}{\sqrt{D}}\right) +%
\frac{\partial}{\partial
v}\left(\frac{n\mathrm{v}_{v}}{\sqrt{D}}\right)=0, \label{nu}
\end{equation}
where
\begin{eqnarray}
f_{1}(u,v) &=&\frac{\partial \mathbf{e}_{u}}{\partial u}\cdot \mathbf{e}_{v}%
\text{ \ , \ \ }f_{2}(u,v)=\frac{\partial \mathbf{e}_{u}}{\partial
v}\cdot
\mathbf{e}_{v}  \nonumber \\
f_{3}(u,v) &=&\frac{\partial \mathbf{e}_{v}}{\partial u}\cdot \mathbf{e}_{u}%
\text{ \ , \ \ }f_{4}(u,v)=\frac{\partial \mathbf{e}_{v}}{\partial
v}\cdot \mathbf{e}_{u}  \label{3.11}
\end{eqnarray}
Now let us consider the case of ''coordinate flows'', when the
flow is directed along one of the families of coordinate lines
$u,v$ for example along $u$-coordinate lines and is given by
$\mathbf{v}=(0,\mathrm{v}_{v}(u,v),0)$, the density
distribution is $n=n(u,v)$. The case of a velocity field $\mathbf{v}=\mathrm{v}%
_{u}\mathbf{e}_{u}$ is equivalent (just interchange $u$ and $v$).
From Eq.~\eqref{vu1} we have:
\begin{equation}\label{vv}
  \mathrm{v}_{v}= C\exp \left[ \int
I(u,v)du\right]\,,
\end{equation}
where
\begin{equation}\label{Iuv}
  I(u,v) =
  \f{1}{2}\,\frac{\partial\,\ln{D}}{\partial\,u}
  - \f{f_4}{s_1\,n} - \f{s_2}{s_1}\frac{\partial\,\ln n}{\partial\,u}\,.
\end{equation}
Equations \eqref{vv1} and \eqref{nu} take the form:
\begin{eqnarray}
\frac{\partial\mathrm{v}_{v}}{\partial v} &=&0,  \label{3.12a} \\
\frac{\partial }{\partial v}\left(
\frac{n\mathrm{v}_{v}}{\sqrt{D}}\right) &=&0 \label{3.12c}
\end{eqnarray}
and lead to
\begin{equation} n(u,v)=h(u)\sqrt{D(u,v)},
\label{3.15}
\end{equation}
where $h(u)$ is an arbitrary function of $u$. Taking into account
that:
\begin{equation}
f_{4}(u,v)=\left( \frac{\partial \mathbf{e}_{v}}{\partial v}\cdot
\mathbf{e}
_{u}\right) =-\frac{D}{2}\frac{\partial \left( 1/D\right) }{\partial u%
}=\frac{1}{2}\frac{\partial \ln D}{\partial u},  \label{3.17}
\end{equation}
and Eq.~\eqref{3.15} we obtain:
\begin{equation}\label{Iuv1}
  I(u,v) =
  \f{1}{2}\,\frac{\partial\,\ln{D}}{\partial\,u}\left[
  1- \f{1}{s_1\,h(u)\sqrt{D}} - \f{s_2}{s_1}\left(2\frac{\partial\, \ln{h(u)}}{\partial\,u}
  \left(\frac{\partial\,\ln{D}}{\partial\,u}\right)^{-1}+1\right)\right]\,.
\end{equation}
Note that as it follows from Eq.~\eqref{3.12a}
\begin{equation}
\mathrm{v}_{v}(u,v)=\mathrm{v}_{v}(u).  \label{3.13}
\end{equation}
For the integrand in Eq.(\ref{vv}) this implies that
\begin{equation}
I(u,v)=I(u).  \label{3.14}
\end{equation}
Therefore from Eqs.~\eqref{Iuv1} and \eqref{3.14} we can conclude
that the function $D(u,v)$, which determines the coordinate
system, depends only on one variable, $D=D\,(u)$.

In the case of conformal coordinates, defined by the metrics in
Eq.~\eqref{3.3}, the Gaussian curvature of the surface is given by
\cite{dubrovinovikov}:
\begin{equation}
K=\frac{1}{2}D\Delta\ln{D}\,.  \label{3.19}
\end{equation}
For a planar flows the condition $K=0$ leads to the following
\begin{equation}
\Delta \ln D=0.  \label{3.21}
\end{equation}
Using the expression for the Laplacian in the conformal
representation (see Eqs.~\eqref{3.5d}) and taking into account a
fact that $D=D(u)$ one finds for \eqref{3.21}:
\begin{equation}
D=\exp\left[c_{1}u+c_{2}\right], \label{3.22}
\end{equation}
where $c_{1,2}$ are arbitrary constants. The case $c_1=0$
determines a Cartesian coordinate system, which related to a
linear class of stationary flow. The case $c_1\ne 0$ determines a
polar coordinate system \cite{madelung}, which corresponds to an
axially symmetric (or vortical) type of flow.

Finally the velocity field for the LHM, with $s_1\ne 0$ takes the
form:
\begin{equation}
\mathrm{v}_{v}(u) =C\exp\left[\f{1}{2}\,\int
\frac{\partial\,\ln{D}}{\partial\,u}\left[
  1- \f{1}{s_1\,h(u)\sqrt{D}} - \f{s_2}{s_1}\left(2\frac{\partial\, \ln{h(u)}}{\partial\,u}
  \left(\frac{\partial\,\ln{D}}{\partial\,u}\right)^{-1}+1\right)\right]\,du\right]\,,
  \label{vvfin}
\end{equation}
Thus it is proved that for the case $s_1\ne 0$ the only stationary
solutions are those either with planar or axial symmetry of the
flow.

The case $s_1=0$ (the LM2) is specific because as it follows from
Eqs.~\eqref{vu1} and \eqref{vv1} the velocity field $v_{v}(u)$ is
arbitrary while the density takes the form:
\begin{equation}\label{ns2}
  n = - \f{1}{2\,s_2}\, \ln{D} + n_0\,.
\end{equation}
The statement about the symmetry of the stationary solutions for
such a model is the same as that proved above for the case $s_1\ne
0$. Note that the parameter $\lambda = s_2/s_1$ can be considered
as the weight factor of the rotor chemotaxis contribution.

\section{The properties of radially symmetric stationary solutions}
In this section we investigate the stationary radially symmetrical
solutions for different cases of the local hydrodynamical models.
In our first paper we considered the $s_{2}=0$ case, which we
called the local model one (LM1). Other models correspond to cases
$s_1=0$ and $s_1=s_2=s$ which we will call the LM2 and the LM12
respectively. We consider the finite and infinite flocking
stationary states for these models. It is natural to differentiate
between these two cases by the finiteness of two integrals of
motion - the total number of particles Eq.~\eqref{totnum} and the
kinetic energy Eq.~\eqref{totenrg}. The infinite flocking is
associated with $N$ infinite but finite $T$, while finite flocking
naturally corresponds to both $N$ and $T$ finite. Note that in the
finite flocking behavior one may consider two cases with respect
to the compactness of the $n(r)$. Compactness means that the
density has some upper cut-off beyond which it can be put zero.

\subsection{The properties of the stationary solutions of LM1}
In our previous paper \cite{usepll2005} we considered the
stationary solutions in LM1 and obtained the for velocity field
profile
\begin{equation}
\mathrm{v}_{\varphi }(r)=\frac{C_{1}}{2\pi r}\exp \left( \frac{1}{s_{1}}%
\int\limits_{r_{0}}^{r}\,\frac{1}{r\,n(r^{\prime })}\,dr^{\prime
}\right)\,. \label{2.1}
\end{equation}
Here $r$ is the radial coordinate and $\varphi$ is the angular
one, $r_{0}$ is some radius, which for vortex-like solutions plays
the role of lower cut-off radius of the vortex and determines its
core.

For vortex-like solutions the constant $C_{1}$ in Eq.(\ref{2.1})
is determined by the circulation of the core
\begin{equation}
\oint\limits_{r=r_{0}}\mathbf{v}\cdot d\mathbf{l}=C_{1}.
\label{2.2}
\end{equation}
The spatial character of the solution given in Eq.(\ref{2.1})
strongly depends on the sign of the parameter $s_{1}$. The
finiteness of integrals of motion Eq.~\eqref{totnum} and
Eq.~\eqref{kin} is guaranteed by either the fast enough decrease
of the density $n(r)$ at $r\to \infty$ or its compactness ($n(r)$
as a function has finite support).

Let us consider the finite flocking behavior (FFB), which is
characterized by both $N$ and $T$ finite. If at $r\to \infty$
asymptotically $n(r)\sim r^{-\alpha }$ , where $\alpha
> 2$ the total number of particles $N$ is finite. Then
at such a behavior of $n(r)$ the total kinetic energy is finite
only if $s_1 < 0$.

\begin{figure}[!hbt]
\centering
\includegraphics[angle=-90,scale=0.28]{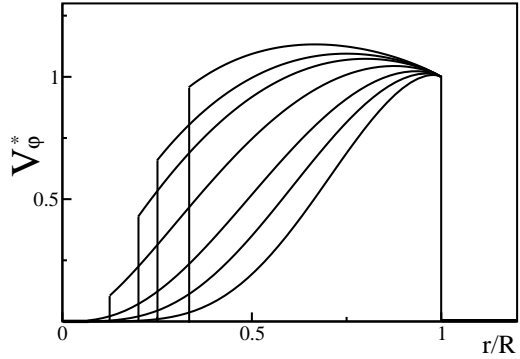}
\caption{The velocity profiles for ${\rm V}^{*}_{\varphi} = 2\pi R
v_{\varphi} (r/R)$ in the LM1 for different ratios $R/r_0$ and
$s_1 n_1 =1$.} \label{dannye}
\end{figure}

In a case  $s_1 > 0$ the condition of finiteness for the kinetic
energy and the total number of particles is fulfilled only if
$n(r)$ has finite support. As an example we may give:
\begin{equation}
n(r)=n_{1}\theta \left( r-r_{0}\right) \,\theta \left( R-r\right) \sqrt{%
\frac{r_{0}}{R-r}}\,,\quad n_1>0\,,  \label{2.3}
\end{equation}
where $R$ is the upper cut-off radius. Substituting Eq.(\ref{2.3})
into Eq.(\ref{2.1}) one obtains:
\begin{equation}
\mathrm{v}_{\varphi }(r)=\frac{C_{1}}{2\pi r}\exp \left[ \frac{2}{s_{1}n_{1}}%
\sqrt{\frac{R}{r_{0}}}\left. \left( \sqrt{1-\frac{r}{R}}-\mathop{\rm arctanh}%
\nolimits\sqrt{1-\frac{r}{R}}\right) \right| _{r_{0}}^{r}\right].
\label{2.4}
\end{equation}
The corresponding profiles for the velocity
$\mathbf{v}=\mathrm{v}_{\phi }\, \mathbf{e}_{\phi }$ at different
ratio $R/r_0$ are shown in Fig.~\ref{dannye}. Note that for the
case considered at $R/r_0 \ge 2$ we get the monotonic profiles of
the velocity which are similar to those observed in experiments
\cite{cvbenjacohenpre1996}.

The infinite flocking behavior (IFB) is characterized by the $N$
infinite and $T$ finite. For the case $s_1>0$ no physical
solutions exist with such a behavior. For the case $s_1 < 0$
slowly decaying density distributions $n(r)\propto r^{-\alpha}$ at
$r \to \infty$ with $0\le \alpha \le 2$ are consistent with the
finiteness of $T$. These statements are summed up in
Table~\ref{tab1}.
\begin{table}[!hbt]
\begin{tabular}{|c|c|c|}
\hline
LM1 & $s_{1}>0$ & $s_{1}<0$ \\
\hline
FFB $\left(N<\infty,\,T<\infty\right)$ & compact support & $\alpha>2$, no compact support \\
\hline
IFB $\left(N=\infty,\,T<\infty\right)$ & no physical solutions & $0\le\alpha\le2$ \\
\hline
\end{tabular}
\caption{Properties of the stationary solutions of the LM1.}
\label{tab1}
\end{table}

\subsection{The properties of the stationary solutions of LM2}
For the case that $s_{1}=0$\ and $s_{2}$ finite one may also
construct a radially symmetric stationary planar solution. In
polar coordinates from Eq.~\eqref{ns2} we get:
\begin{equation}
n(r)=\frac{1}{s_{2}}\ln \frac{r}{r_{0}}\,,  \label{2.7}
\end{equation}
and one can choose the velocity field $\mathrm{v}_{\varphi }(r)$\
arbitrarily. For positive values of $s_{2}$\ this density is
positive for $r>r_{0}$\ and for negative values of $s_{2}$\ it is
positive for $r<r_{0}$. So for positive values of $s_{2}$\ the
density profile becomes
\begin{equation}
n(r)=\frac{1}{s_{2}}\,\theta \,(R-r)\,\ln \frac{r}{r_{0}}
\label{2.8}
\end{equation}
and for negative values of $s_{2}$ it becomes
\begin{equation}
n(r)=\frac{1}{\left| s_{2}\right| }\,\theta \,(r_{0}-r)\ln
\frac{r_{0}}{r}.  \label{2.9}
\end{equation}
The results about finite and infinite flocking behavior are in
Table~\ref{tab2}.
\begin{table}[!hbt]
\begin{tabular}{|c|c|c|}
\hline
LM2 & \,$s_{1}>0\,$ &\, $s_{1}<0$\, \\
\hline
FFB $\left(N<\infty,\,T<\infty\right)$\, &\, no physical solution\, &\, compact support \\
\hline
IFB $\left(N=\infty,\,T<\infty\right)$\, & \,no physical solution \,&\, no physical solution \\
\hline
\end{tabular}\caption{Properties of the stationary solutions of the LM2.} \label{tab2}
\end{table}

\subsection{The properties of the stationary solutions of LM12}
The third case which is expedient to consider is $s_{1}=s_{2}=s$.
In that case, according to Eq.~\eqref{lhm}, the field
$\bs{\omega}$ is coupled to the number density flux $\m{j} = n
\,\mathbf{v}$:
\begin{equation}
\boldsymbol{\omega }\left( \mathbf{r},t\right) = s\,\,\mathop{\rm
rot} \m{j}   \label{2.11}
\end{equation}
so that Eq.~\eqref{lhmv} for the velocity is
\begin{equation}
\frac{d}{dt}\mathbf{v}\left( \mathbf{r},t\right) =s\,\mathop{\rm
rot}\m{j} \times \mathbf{v}\left( \mathbf{r},t\right)\,.
\label{2.12}
\end{equation}
For a radially symmetric stationary planar
solution this gives
\begin{equation}
\mathrm{v}_{\varphi }(r)=s\frac{d}{dr}\left[ rn(r)\mathrm{v}_{\varphi }(r)%
\right] .  \label{2.13}
\end{equation}
with
\begin{equation}
\mathrm{v}_{\varphi }(r)=\frac{C}{2\pi r\,n(r)}\exp \left[ \frac{1}{s}%
\int_{r_{0}}^{r}\frac{1}{r^{\prime }\,n\,(r^{\prime
})}\,dr^{\prime }\right]\,.  \label{2.14}
\end{equation}
as a solution. The constant $C$ is determined by the circulation
of the core
\begin{equation}
\oint\limits_{r=r_{0}}\mathbf{v}\cdot
d\mathbf{l}=\frac{C}{n(r_{0})}\,. \label{2.15}
\end{equation}
The properties of finite and infinite flocking behavior for this
model are the same as those for the LM1 (see Table~\ref{tab1}).

\section{Conclusions}
In this paper we consider the properties of the stationary 2D
solutions of the LHM proposed in \cite{usepll2005}. We established
that the only possible stationary solutions in the model are those
with translational or axial symmetry. The cases of finite and
infinite flocking behavior are considered for different specific
types of the LHM. It is shown that the case $s_1=0$ (LM2) is
specific in a sense that there is only one density distribution,
for which many velocity profiles can be realized. In general case
($s_1\ne 0$) one is free to choose axially symmetric density
distribution which the velocity profile depends on
(Eqs.~\eqref{3.15} and \eqref{vvfin}). Note that in this respect
the general case is similar to the LM1 considered earlier.

\begin{ack} Vladimir Kulinskii thanks the
Nederlandse Organisatie voor Wetenschappelijk Onderzoek (NWO) for
a grant, which enabled him to visit Dick Bedeaux's group at Leiden
University.
\end{ack}

%\bibliography{sppbib}

\end{document}